\documentclass{emulateapj}
\usepackage{graphicx,amsmath,amssymb}
\usepackage{apjfonts}
\usepackage{natbib}
\bibpunct[, ]{(}{)}{;}{a}{}{,}

\begin{document}

   \title{A new cosmological distance measure using AGN}

   \author{D.~Watson, K.~D.~Denney, M.~Vestergaard}
   \affil{Dark Cosmology Centre, Niels Bohr Institute, University of Copenhagen, Juliane Maries Vej 30, DK-2100 Copenhagen \O, Denmark\\ \texttt{darach@dark-cosmology.dk, kelly@dark-cosmology.dk, vester@dark-cosmology.dk}}
   \and
   \author{T.~M.~Davis}
   \affil{School of Mathematics and Physics, University of Queensland, Brisbane, QLD 4072, Australia\\ \texttt{tamarad@physics.uq.edu.au}}

   \begin{abstract}

Accurate distances to celestial objects are key to establishing the age and
energy density of the Universe and the nature of dark energy.  A distance
measure using active galactic nuclei (AGN) has been sought for more than
forty years, as they are extremely luminous and can be observed at very
large distances.  We report here the discovery of an accurate luminosity
distance measure using AGN.  We use the tight relationship between the
luminosity of an AGN and the radius of its broad line region established via
reverberation mapping to determine the luminosity distances to a sample of
38 AGN.  All reliable distance measures up to now have been limited to
moderate redshift---AGN will, for the first time, allow distances to be
estimated to $z\sim4$, where variations of dark energy and alternate gravity
theories can be probed.

   \end{abstract}

   \maketitle

%
%
\section{Introduction\label{introduction}}

One of the simplest and, perversely, most intractable problems in astronomy
has been to discover how far away something is.  New distance measures have
led to fundamental changes in our understanding of the Universe; for example
Tycho Brahe's supernova and Edwin Hubble's Cepheids radically reshaped our
understanding of the cosmos.  It is almost two decades since type Ia
supernovae (SNe) were shown to be accurate standard candles
\citep{1993ApJ...413L.105P}.  That distance measure led directly to the
discovery of the acceleration of the Universe and dark energy
\citep{1998AJ....116.1009R,1999ApJ...517..565P}.  Finding reliable methods
to determine distances, especially large distances in the Hubble flow, is an
ongoing task.  In particular, reliable distances beyond redshift $z\sim1.7$
are beyond the scope of current tools.  Investigating the evolution of the
dark energy equation of state has therefore been very limited up to now. 
Since their discovery over four decades ago it has been hoped that quasars,
or the more general class of active galactic nuclei (AGN) of which quasars
are a subset, could be used as standard candles for cosmology.  Many
attempts were made
\citep{1977ApJ...214..679B,1999MNRAS.302L..24C,2002ApJ...581L..67E,2003MSAIS...3..218M},
none very successful.  However, recent advances in our understanding of AGN,
which are extremely luminous, common, and readily observable over a range of
distances from $\sim10$\,Mpc to $z>7$, prompted us to investigate their use
as standard candles for cosmology.

Here we examine the possibility of using the relationship
between the luminosities of type~1 AGN and the sizes of their broad-line
regions established via reverberation mapping as a luminosity distance
indicator.  In section~\ref{method} we describe our methods, in
section~\ref{observations} we detail the data used.  In
section~\ref{results} we provide the results of our analysis and the sources
of scatter in the relation.  Section~\ref{discussion} contains a discussion
on the prospects for this new distance measure, its competitiveness and
unique features. Uncertainties quoted are at the 68\%
confidence level.  A
cosmology where $H_0=70.2$\,km\,s$^{-1}$\,Mpc$^{-1}$, $\Omega_\Lambda =
0.725$ and $\Omega_{\rm m}=0.275$ is assumed throughout.

%
%
\section{Method}\label{method} 

\subsection{Broad-line region reverberation mapping} 

The supermassive black hole that lies at the heart of every AGN and is its
ultimate power source, is surrounded at a distance by high velocity gas
clouds that produce the broad emission lines characteristic of the spectra
of near--face-on AGN, i.e.\ quasars and Seyfert~1 galaxies.  It has been
known for some time that there is a relationship between the size of this
broad-line emitting region (BLR) and the AGN's central continuum luminosity
\citep{2000ApJ...533..631K,2005ApJ...629...61K,2009ApJ...697..160B}.  The
size of the BLR is determined by the depth to which the surrounding
gas can be photo-ionised by the central source.  Since the ionising flux
drops with distance according to the inverse square law, the radius of the
BLR, $r$, is expected to be proportional to the square root of
the luminosity, $\sqrt{L}$ \citep{1975ApJ...202..306M}.  Establishing $r$
and the flux would clearly lead to a measure of the luminosity distance to
the source.

The photons emitted by BLR gas are reprocessed continuum photons. 
Therefore, the flux in the broad lines varies in response to variations in
the luminosity of the central source with a time-delay, $\tau$, governed
simply by the light travel time, $\tau = r/c$.  Measuring the time delay
thus allows a determination of the BLR radius, a technique known as
`reverberation mapping' \citep{1982ApJ...255..419B,1993PASP..105..247P}. 
The radius is effectively determined by measuring the time lag between
changes in the continuum luminosity of the AGN and the luminosity of a
bright emission line (typically H$\beta$ or \ion{C}{4}).  The time lag
should therefore be proportional to the square root of the luminosity of the
central source: $\tau\propto\sqrt{L}$.  The observable quantity,
$\tau/\sqrt{F}$, where $\tau$ is the emission line lag and $F$ is the
measured AGN continuum flux, is then a measure of the luminosity distance to
the source.

Recent improvements in lag and luminosity measurements, notably, improving
the luminosities by removing the contaminating effects of the host galaxy
\citep{2009ApJ...697..160B}, improving lag measurements by reobserving AGN
that previously had poorly-sampled lightcurves \citep{2010ApJ...721..715D},
and populating the low luminosity regime of the existing sample
\citep{2009ApJ...705..199B}, have shown that the radius-luminosity
relationship follows the expected $r\propto\sqrt{L}$ relation to good
accuracy across four orders of magnitude in luminosity
\citep{2009ApJ...697..160B,2011ApJ...735...80Z}.  We have used this
relationship to turn a sample of AGN with well-determined lags into standard
candles for cosmology.

\section{Observational data}\label{observations} 
\begin{figure}
 \includegraphics[angle=0,viewport=8.400000 25.000000 404.850000 407.850000,clip=,width=\columnwidth]{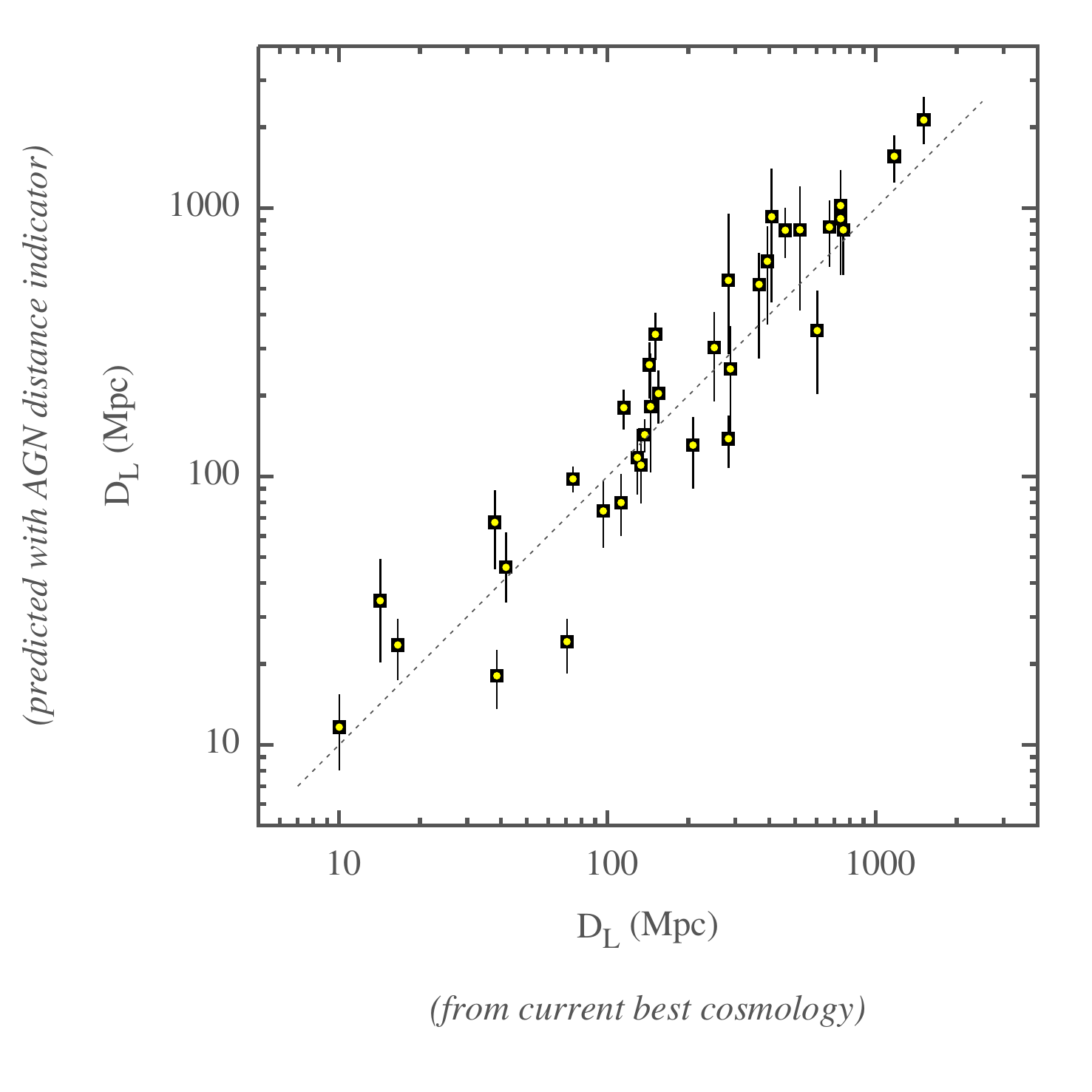}
 \caption{Comparison of AGN-derived distances to Hubble distances based on
          the best current cosmology \citep{2011ApJS..192...18K}. The dotted
          line is the equality of both distances.  The AGN distance
          estimates follow the best current cosmology Hubble distances to
          good accuracy}
 \label{fig:DL_DL}
\end{figure}

To construct our sample of
$\tau/\sqrt{F}$ values, we use all available lags for the H$\beta$ line and
restframe 5100\,\AA\ continuum fluxes
\citep{2009ApJ...697..160B,2010ApJ...721..715D}, where these measurements
have the constant host galaxy component removed from the measured flux to
obtain the AGN continuum flux.  This host galaxy removal has been shown to
be very important \citep{2009ApJ...697..160B} and is done primarily using
images from the \emph{Hubble Space Telescope} (\emph{HST}) to model the
underlying galaxy contribution.  We correct the fluxes for Galactic
extinction \citep{1998ApJ...500..525S,2010ApJ...725.1175S}.  We do not apply a
correction for internal extinction because extinction
estimates have been made for only a fraction of the objects in our sample
and typically the discrepancy between estimates in a single object are as
large as the extinction correction itself (see below for a more detailed
discussion of intrinsic extinction correction).  We then calibrate
$\tau/\sqrt{F}$ to the absolute distance for the source NGC\,3227
\citep{2001ApJ...546..681T,2004AJ....128.3034K}.  Fig.~\ref{fig:DL_DL} shows
that the luminosity distances we determine for our sample closely
follow the predicted distances for the current best-estimated
\emph{WMAP}-$\Lambda$CDM cosmology \citep{2011ApJS..192...18K}.

\subsection{The absolute distance calibration}
Currently only NGC\,3227 and NGC\,4051 have direct distance estimates.  The
Tully-Fisher distance to NGC\,4051 is the less accurate of the two.  We
therefore calibrated $\tau/\sqrt{F}$ to the luminosity distance to the
galaxy NGC\,3227 based on a distance modulus of $m-M=31.86\pm0.24$
determined using the surface brightness fluctuations (SBF) method to its
companion galaxy NGC\,3226 \citep{2001ApJ...546..681T}.  Due to the
occurrence of a supernova, 2002bo, in another member of the Leo\,III group
to which NGC\,3227 belongs, the SBF distance quoted above has been examined
in detail and seems likely to be correct within the quoted uncertainty
\citep{2004AJ....128.3034K}.  However, the uncertainty in this calibration
is relatively large, and we use it here only to determine an initial
estimate of the luminosity distances.  NGC\,4051, NGC\,4151, and NGC\,3227
are certainly close enough that it should be possible to obtain more
reliable Cepheid-derived distances with \emph{HST} to these galaxies for a
better absolute calibration.  In practice, we expect that Cepheid distances
can in fact be determined to multiple nearby AGN, allowing at least a dozen
AGN to be distance-calibrated in this way.

%
%
\section{Results}\label{results}

\subsection{How tight is the relation?}

In order to be a successful distance indicator, $\tau/\sqrt{F}$ must follow
the luminosity distance with little inherent scatter. We estimate below the
different sources of scatter in the relation with the current data and
determine how far this scatter can be reduced in the immediately foreseeable
future. This is summarised in Table~\ref{tab:scatter}.

\begin{table}
\caption{Scatter in the AGN Hubble diagram}
\label{tab:scatter}
\setlength{\tabcolsep}{3pt}
 \begin{center}
  \begin{tabular}{@{}lcc@{}}
   \hline\hline
    Source of scatter\footnote{Root mean square scatter in dex (mag)} & Current & Can be reduced to \\
   \hline
    Observational & 0.14 (0.36) & 0.05 (0.13)\\
    Extinction & 0.08 (0.20) & 0.04 (0.10)\\
    Bad lags & 0.11 (0.28) & 0.00 (0.00)\\
    Other & 0.05 (0.13) & 0.05 (0.13)\\
   \hline
    Total & 0.20 (0.50) & 0.08 (0.20)\\
   \hline
  \end{tabular}
 \end{center}
\end{table}

\begin{figure*}
\begin{centering}
 \includegraphics[angle=0,viewport=38.200000 64.100000 626.400000 598.350000,clip=,width=0.75\textwidth]{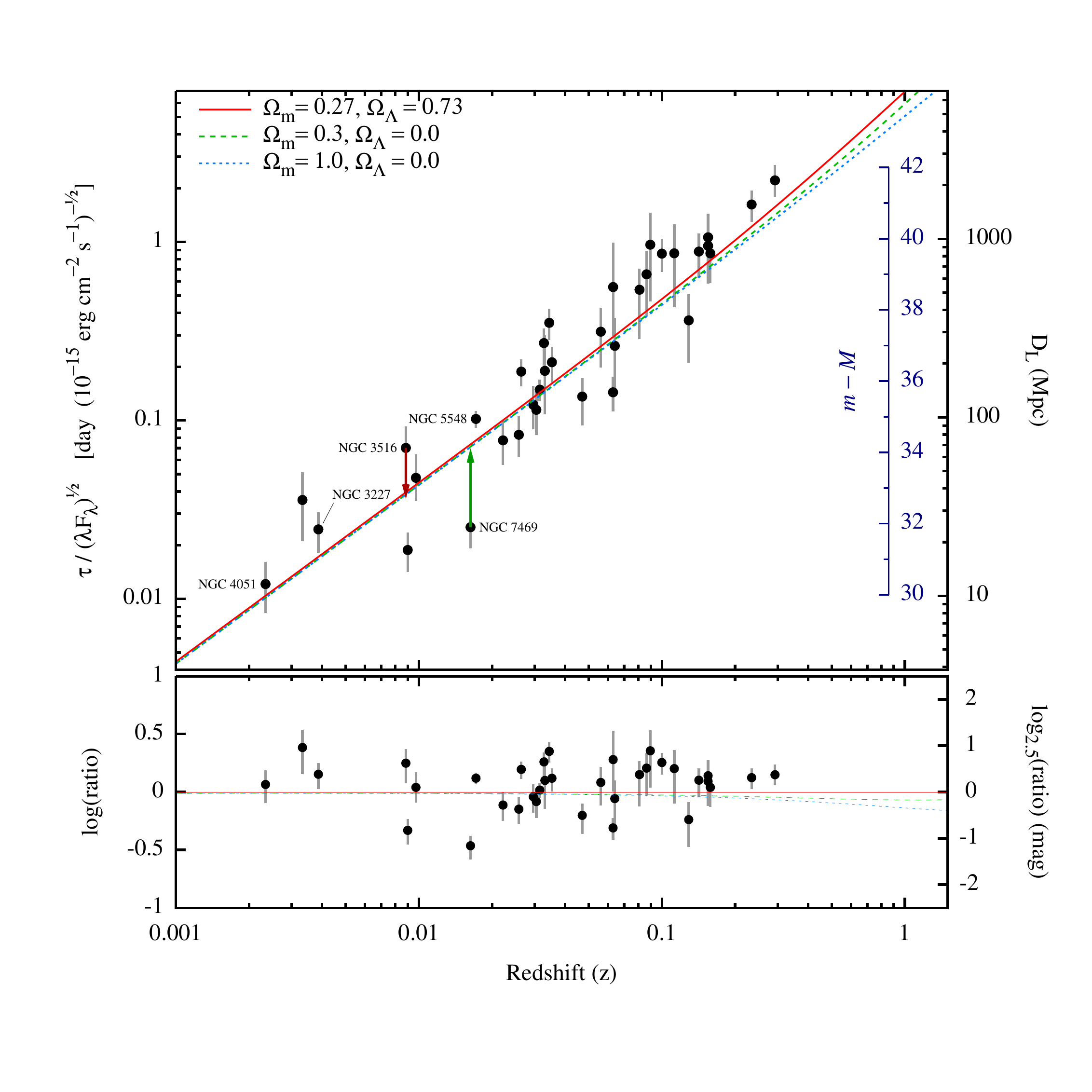}
 \caption{The AGN Hubble diagram. The luminosity distance indicator
          $\tau/\sqrt{F}$ is plotted as a function of redshift for 38 AGN
          with H$\beta$ lag measurements.  On the right axis the luminosity
          distance and distance modulus ($m-M$) are shown using the surface
          brightness fluctuations distance to NGC\,3227 as a calibrator. 
          The current best cosmology \citep{2011ApJS..192...18K} is plotted
          as a solid line.  The line is not fit to the data but clearly
          follows the data well.  Cosmologies with no dark energy components
          are plotted as dashed and dotted lines.  The lower panel shows the
          logarithm of the ratio of the data compared to the current
          cosmology on the left axis, with the same values but in magnitudes
          on the right.  The red arrow indicates the correction for internal
          extinction for NGC\,3516.  The green arrow shows where NGC\,7469
          would lie using the revised lag estimate from
          \citet{2011ApJ...735...80Z}.  NGC\,7469 is our largest outlier and
          is believed to be an example of an object with a misidentified lag
          \citep{2010IAUS..267..151P}.}
 \label{fig:Hubble_diagram}
\end{centering}
\end{figure*}

We have estimated the
root mean square scatter in our AGN Hubble diagram
(Fig.~\ref{fig:Hubble_diagram}) to be 0.2\,dex, equivalent to 0.5\,mag in
distance modulus.  Based on the expectation of a reduced $\chi^2$ value of
unity, we have determined the observational uncertainty to account for
$\sim50\%$ of the total scatter in the relation, or 0.14\,dex (0.36\,mag).

\subsubsection{Scatter due to observational uncertainty}
\begin{figure}
 \includegraphics[angle=0,viewport=2.500000 7.0 396.95 270.2,clip=,width=\columnwidth]{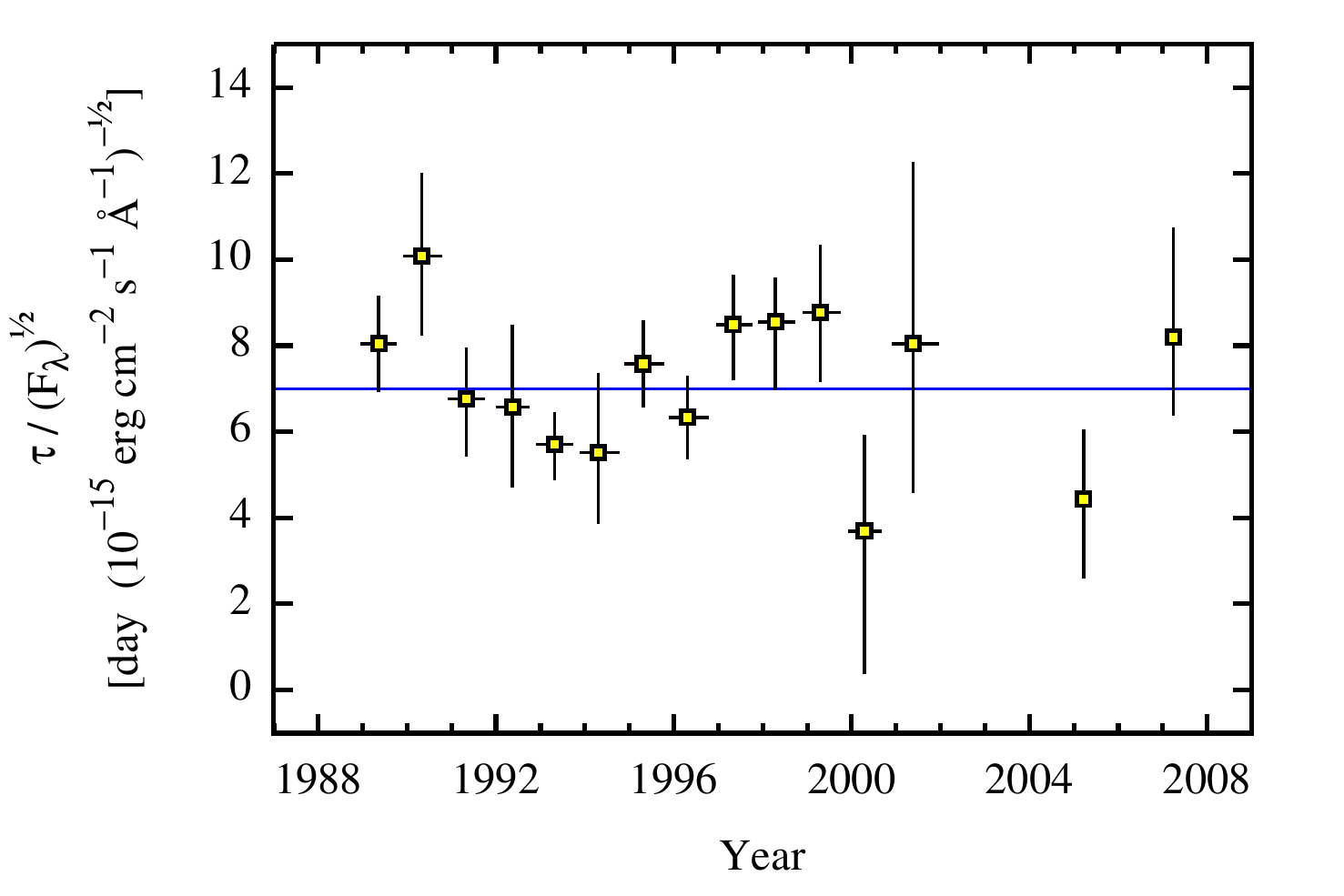}
 \caption{The distance indicator for NGC\,5548, the best observed AGN in our
          sample, showing that the indicator is constant over time. The
          scatter in the data is consistent with the statistical
          uncertainty, showing that the scatter in excess of the
          observational uncertainty in the AGN Hubble diagram is largely due
          to object-to-object variations}
 \label{fig:NGC5548_trootF}
\end{figure}

The scatter due to observational uncertainty can be reduced significantly. A
major advantage held by AGN is that they can be observed repeatedly and the
distance to any given object substantially refined.  For example, the
observational uncertainty for NGC\,5548 is 0.05\,dex (0.13\,mag) after
roughly a dozen reverberation measurements, while it is typically
$\sim0.14$\,dex (0.35\,mag) for sources with single measurements.  In
Fig.~\ref{fig:NGC5548_trootF} we show $\tau/\sqrt{F}$ from all observing
campaigns of NGC\,5548.  The reduced $\chi^2$ is very close to unity for a
constant value fit to the data, indicating that the scatter is almost
entirely accounted for by the uncertainty related to the observations. 
There is thus very little intrinsic variation in $\tau/\sqrt{F}$ for a given
object.  This means that repeated observations of given sources will reduce
the scatter related to observational uncertainty to a level similar to that
observed in NGC\,5548.

\subsubsection{Extinction}

Another likely source of scatter is due to extinction associated with the
AGN and its host galaxy.  As an example, in Fig.~\ref{fig:Hubble_diagram} we
plot the shift in $\tau/\sqrt{F}$ resulting from the extinction correction
for NGC\,3516 \citep{2010ApJ...721..715D}.  This shift moves NGC\,3516 very
close to the best-fit line.  Currently, few reliable extinction measurements
exist for these AGN.  We show the extinction correction for NGC\,3516, in
Fig.~\ref{fig:Hubble_diagram}, however, because it may be reliable as
it is based on two different extinction estimations with different methods
that give consistent results \citep{2010ApJ...721..715D}.  An accurate
correction for internal extinction should reduce the scatter in the Hubble
diagram.  A mean scatter of $\sim0.16$\,dex (0.4\,mag) as suggested
by \citet{2007MNRAS.380..669C}, induced in the luminosities of these AGN by
internal extinction, would contribute 0.08\,dex (0.2\,mag) scatter to the
relationship.  Using the Balmer decrement method, \ion{Na}{1}\,D or
\ion{K}{1} line equivalent widths, or a calibration based on several
methods, we would then expect to be able to reduce the scatter induced by
extinction from 0.08\,dex (0.2\,mag) to the level at which these
correlations are calibrated, 0.04\,dex (0.1\,mag)
\citep{1997A&A...318..269M}.

Adding an extinction correction will of course systematically make the
distances smaller, but since the objects are currently calibrated to the
distance to NGC\,3227, it is the extinction relative to this object that
matters.  Currently we assume that the extinction of NGC\,3227 is reasonably
close to the mean extinction of the sample.  The fact that the distances are
are not noticeably offset from the distances based on current best
estimates of $H_0$ (Fig.~\ref{fig:DL_DL}) suggests that this is a
reasonable assumption.

\subsubsection{Incorrect Lags} 

The scatter due to an apparent diversity between objects can also be rapidly
reduced.  While we use all available H$\beta$ lag detections with published
host-corrected AGN fluxes here to avoid biasing our results, it has been
shown \citep{2010IAUS..267..151P} that with selection of only good quality
lag measurements, the scatter in the observed radius-luminosity relation can
be reduced to a level almost consistent with the observational uncertainty. 
This suggests that some of the lag measurements currently in use are
incorrect at a level much larger than their quoted uncertainties.  This may
in part be due to developing observational expertise, where later lag
measurements benefitted from improved observing practices, e.g.\
better-sampled lightcurves.  NGC\,7469 alone contributes 0.09\,dex
(0.22\,mag) to the total scatter.  The lags of most of our objects were
recalculated by \citet{2011ApJ...735...80Z} using the stochastic process
estimation for AGN reverberations (SPEAR) method.  The only object with a
significantly discrepant lag they found was NGC\,7469.  A more recent
observation of this source (C.~Grier, private communication) has also given
a value of $\tau/\sqrt{F}$ consistent with the SPEAR value
\citep{2011ApJ...735...80Z}.  Using the SPEAR value for its lag (see
Fig.~\ref{fig:Hubble_diagram}) the scatter in the diagram is reduced to
0.18\,dex (0.45\,mag), and the reduced $\chi^2$ drops from 4.0 to 2.9.  A
small number of misidentified lags can contribute a very large fraction of
the scatter in excess of the observational uncertainty and should easily be
spotted and removed with multiple observations of the same source.

We have noted above how an observational error contribution to the total
scatter of $\sim0.05$\,dex (0.12\,mag) or less is achievable with multiple
observations of each source.  In addition, correcting NGC\,7469 alone is
enough to reduce the component of the scatter unrelated to observational
uncertainty from 0.14\,dex (0.36\,mag) to 0.11\,dex (0.28\,mag).  Choosing
only good lags improves the scatter in excess of the observational
uncertainty to a level consistent with the extinction-related scatter
estimated above \citep{2010IAUS..267..151P}.  With many more good quality
observations it therefore seems likely that the scatter unrelated to
observational uncertainty will be reduced to a level dominated by the
uncertainty in the extinction correction, $\sim0.04$\,dex (0.1\,mag, see
above).  With these improvements, the total known scatter in the
relationship could therefore be reduced to 0.06\,dex (0.16\,mag), dominated
by the extinction correction and observational uncertainties (see
Table~\ref{tab:scatter}).

Given these straightforward improvements in the statistical quality of the
measurements via an increased number of observations per source, more
reliable lags, and better extinction estimates, the scatter in the AGN
Hubble diagram can be decreased very substantially with medium-sized
programmes on existing facilities in only a few years.  Based on the
arguments made above we estimate that the total scatter in the relationship
could, in a few years observing, be reduced to 0.08\,dex (0.20\,mag, see
Table~\ref{tab:scatter}), not far in excess of the current SNIa
samples \citep{2009ApJS..185...32K,2011ApJS..192....1C}.

%
%

\section{Discussion}\label{discussion}

\subsection{Prospects for extension to high redshift} 
The unique aspect of the AGN Hubble diagram is that while SN distances are
currently observationally restricted to less than $z\sim1.7$
\citep{2001ApJ...560...49R}, and are unlikely to exceed $z=2$, there is no
such impediment to extending the AGN diagram to the redshift range
$z=0.3-4$, where the power to discriminate dark energy models lies.  Effects
related to changes with time in the equation of state of dark energy or to
alternate gravity theories should be more apparent with higher redshift data
and can only be constrained very poorly with current methods.  We already
know that the radius-luminosity relationship holds in an unaltered form over
four decades in luminosity
\citep{2000ApJ...533..631K,2009ApJ...697..160B,2011ApJ...735...80Z}.  We
have no reason to believe that it should change with redshift either, as its
form is based strictly on the simplest photo-ionization physics.

In practice, extending the diagram to high redshifts requires substantially
longer temporal baselines: 1) redshift increases the observed-frame lags due to time dilation
effects, 2)at higher redshift we observe more luminous AGN,
which have larger BLRs and hence larger rest-frame lags.  For
example, for a redshift $z\sim2$ AGN, we expect a H$\beta$ observed lag of
$\sim2$ years based on the AGN Hubble diagram and assuming an
observed continuum flux of
$1\times10^{17}$\,erg\,cm$^{-2}$\,s$^{-1}$\,\AA$^{-1}$ at a rest-frame
wavelength of 5100\,\AA.  For H$\beta$ the likely minimum time required to
obtain a lag reaches close to a decade at $z\sim2.5$, making this the
practical redshift limit for obtaining lags with H$\beta$.  On the other
hand, strong UV lines such as \ion{C}{4} 1549\,\AA\ are readily observable even for the
most distant known AGN.  Since the BLR is
ionization-stratified \citep{1993ApJ...408..416D}, higher excitation lines
such as \ion{C}{4} have the advantage that they are emitted closer to the
central source and hence have shorter lag times.  And it is worth noting
that a detection of a lag at $z\sim2$ has already been made using
\ion{C}{4} \citep{2007ApJ...659..997K}.  For \ion{C}{4}
, the lags are typically shorter than H$\beta$ by a factor of
2--3 \citep{1991ApJ...366...64C}, meaning that \ion{C}{4} lags should be
measurable for objects up to $z\sim4$.  It has been shown that the
\ion{N}{5} 1240\,\AA\ line has an even shorter lag time than \ion{C}{4}, by
another factor of 2--3 \citep{1991ApJ...366...64C}, allowing lag
measurements to be made in principle to $z\sim6$.  But disentangling the
\ion{N}{5} flux from Ly$\alpha$ may not be straightforward, and we do not
believe that distance measurements are currently feasible with available
facilities beyond $z\sim4$.  Using the \ion{C}{4} line as a lag estimator
has the added advantage that no host galaxy subtraction needs to be made,
thus reducing the flux uncertainty further, somewhat improving our accuracy
at high redshift.  It is worth noting as well that higher redshift sources
require roughly the same total number of observations to obtain a lag
measurment as lower redshift sources, but spread over a longer time.  This
means that, counter-intuitively, high-redshift campaigns are less resource
intensive in a given observing semester.

\subsection{Absolute distances}
The AGN Hubble diagram, like SNe Ia, currently requires calibration to
absolute distances. 
Our developing understanding of the radius-luminosity
relation for the BLR means that direct determination of $H_0$
may be possible with AGN, though via a different avenue than previously
imagined \citep[e.g.][]{1999MNRAS.302L..24C,2007MNRAS.380..669C}.  With direct
observations of the densities and structures of BLRs, the
H$\beta$ or \ion{C}{4} radius may be obtained from first principles,
ultimately obviating the need for the distance ladder in determining
cosmological distances.  While currently we have no means to directly
determine the densities in the BLR, we do have methods
\citep{1994ASPC...69...23H} to begin to dissect the BLR
structure using velocity-resolved time lags \citep{2009ApJ...704L..80D} and
velocity delay maps \citep{2010ApJ...720L..46B}, and future work may reveal
density diagnostics for the BLR.

\subsection{Intrinsic diversity between objects}
The ultimate limit of the accuracy of the method will rely on how the BLR
responds to changes in the luminosity of the central source.  The current
tight radius-luminosity relationship indicates that the ionisation parameter
and the gas density are both close to constant across our sample.  Where the
ionisation parameter is $U\propto L/(n_er^2)$, with $L$ being the luminosity
of the source, $n_e$ the electron density and $r$ the distance between the
source and the gas.  That the ionisation parameter might be constant is not
unexpected based on the locally optimally-emitting cloud model
\citep{1995ApJ...455L.119B}.  However, the mechanism that keeps the density
at close to the same value in a given region of the BLR,
across sources, and for a wide range of luminosities, is unclear.  Precisely
how constant the density is, seems likely to be the factor ultimately
limiting the accuracy of the AGN Hubble diagram if the extinction correction
can be perfected.  Now and for the foreseeable future, however,
observational uncertainty, misidentified lags, and extinction dominate the
scatter in the diagram.

\subsection{Accuracy and competitiveness}
The AGN Hubble diagram seems likely to be competitive with the best current
distance indicator at moderate redshifts within a decade.  The facilities
required up to $z\sim1.5$ are only small ground-based telescopes with
low-resolution spectrographs.  At higher redshifts, where it is the only
distance measure, moderate aperture telescopes and years of monitoring will
be required to obtain lags for a significant number of sources.  However, if
experiments of the scale currently being employed for cosmography are
applied to the AGN Hubble diagram, constraints competitive with the best
current methods should be achievable in less than a decade at redshifts up
to $z=3$.

%
%

\begin{acknowledgements} 

The Dark Cosmology Centre is funded by the DNRF. We would like to thank
Bradley M.  Peterson, Jens Hjorth, Anja C.  Andersen for discussions and
comments on the manuscript.

\end{acknowledgements}


\begin{thebibliography}{36}
\expandafter\ifx\csname natexlab\endcsname\relax\def\natexlab#1{#1}\fi

\bibitem[{{Antonucci} \& {Miller}(1985)}]{1985ApJ...297..621A}
{Antonucci}, R.~R.~J., \& {Miller}, J.~S. 1985, \apj, 297, 621

\bibitem[{{Baldwin} {et~al.}(1995){Baldwin}, {Ferland}, {Korista}, \&
  {Verner}}]{1995ApJ...455L.119B}
{Baldwin}, J., {Ferland}, G., {Korista}, K., \& {Verner}, D. 1995, \apjl, 455,
  L119

\bibitem[{{Baldwin}(1977)}]{1977ApJ...214..679B}
{Baldwin}, J.~A. 1977, \apj, 214, 679

\bibitem[{{Bentz} {et~al.}(2009{\natexlab{a}}){Bentz}, {Peterson}, {Netzer},
  {Pogge}, \& {Vestergaard}}]{2009ApJ...697..160B}
{Bentz}, M.~C., {Peterson}, B.~M., {Netzer}, H., {Pogge}, R.~W., \&
  {Vestergaard}, M. 2009{\natexlab{a}}, \apj, 697, 160

\bibitem[{{Bentz} {et~al.}(2009{\natexlab{b}}){Bentz}, {Walsh}, {Barth},
  {Baliber}, {Bennert}, {Canalizo}, {Filippenko}, {Ganeshalingam}, {Gates},
  {Greene}, {Hidas}, {Hiner}, {Lee}, {Li}, {Malkan}, {Minezaki}, {Sakata},
  {Serduke}, {Silverman}, {Steele}, {Stern}, {Street}, {Thornton}, {Treu},
  {Wang}, {Woo}, \& {Yoshii}}]{2009ApJ...705..199B}
{Bentz}, M.~C., {et~al.} 2009{\natexlab{b}}, \apj, 705, 199

\bibitem[{{Bentz} {et~al.}(2010){Bentz}, {Horne}, {Barth}, {Bennert},
  {Canalizo}, {Filippenko}, {Gates}, {Malkan}, {Minezaki}, {Treu}, {Woo}, \&
  {Walsh}}]{2010ApJ...720L..46B}
---. 2010, \apjl, 720, L46

\bibitem[{{Blandford} \& {McKee}(1982)}]{1982ApJ...255..419B}
{Blandford}, R.~D., \& {McKee}, C.~F. 1982, \apj, 255, 419

\bibitem[{{Braatz} {et~al.}(2010){Braatz}, {Reid}, {Humphreys}, {Henkel},
  {Condon}, \& {Lo}}]{2010ApJ...718..657B}
{Braatz}, J.~A., {Reid}, M.~J., {Humphreys}, E.~M.~L., {Henkel}, C., {Condon},
  J.~J., \& {Lo}, K.~Y. 2010, \apj, 718, 657

\bibitem[{{Cackett} {et~al.}(2007){Cackett}, {Horne}, \&
  {Winkler}}]{2007MNRAS.380..669C}
{Cackett}, E.~M., {Horne}, K., \& {Winkler}, H. 2007, \mnras, 380, 669

\bibitem[{{Clavel} {et~al.}(1991){Clavel}, {Reichert}, {Alloin}, {Crenshaw},
  {Kriss}, {Krolik}, {Malkan}, {Netzer}, {Peterson}, {Wamsteker}, {Altamore},
  {Baribaud}, {Barr}, {Beck}, {Binette}, {Bromage}, {Brosch}, {Diaz},
  {Filippenko}, {Fricke}, {Gaskell}, {Giommi}, {Glass}, {Gondhalekar},
  {Hackney}, {Halpern}, {Hutter}, {Joersaeter}, {Kinney}, {Kollatschny},
  {Koratkar}, {Korista}, {Laor}, {Lasota}, {Leibowitz}, {Maoz}, {Martin},
  {Mazeh}, {Meurs}, {Nair}, {O'Brien}, {Pelat}, {Perez}, {Perola}, {Ptak},
  {Rodriguez-Pascual}, {Rosenblatt}, {Sadun}, {Santos-Lleo}, {Shaw}, {Smith},
  {Stirpe}, {Stoner}, {Sun}, {Ulrich}, {van Groningen}, \&
  {Zheng}}]{1991ApJ...366...64C}
{Clavel}, J., {et~al.} 1991, \apj, 366, 64

\bibitem[{{Collier} {et~al.}(1999){Collier}, {Horne}, {Wanders}, \&
  {Peterson}}]{1999MNRAS.302L..24C}
{Collier}, S., {Horne}, K., {Wanders}, I., \& {Peterson}, B.~M. 1999, \mnras,
  302, L24

\bibitem[{{Conley} {et~al.}(2011){Conley}, {Guy}, {Sullivan}, {Regnault},
  {Astier}, {Balland}, {Basa}, {Carlberg}, {Fouchez}, {Hardin}, {Hook},
  {Howell}, {Pain}, {Palanque-Delabrouille}, {Perrett}, {Pritchet}, {Rich},
  {Ruhlmann-Kleider}, {Balam}, {Baumont}, {Ellis}, {Fabbro}, {Fakhouri},
  {Fourmanoit}, {Gonz{\'a}lez-Gait{\'a}n}, {Graham}, {Hudson}, {Hsiao},
  {Kronborg}, {Lidman}, {Mourao}, {Neill}, {Perlmutter}, {Ripoche}, {Suzuki},
  \& {Walker}}]{2011ApJS..192....1C}
{Conley}, A., {et~al.} 2011, \apjs, 192, 1

\bibitem[{{Denney} {et~al.}(2009){Denney}, {Peterson}, {Pogge}, {Adair},
  {Atlee}, {Au-Yong}, {Bentz}, {Bird}, {Brokofsky}, {Chisholm}, {Comins},
  {Dietrich}, {Doroshenko}, {Eastman}, {Efimov}, {Ewald}, {Ferbey}, {Gaskell},
  {Hedrick}, {Jackson}, {Klimanov}, {Klimek}, {Kruse}, {Lad{\'e}route}, {Lamb},
  {Leighly}, {Minezaki}, {Nazarov}, {Onken}, {Petersen}, {Peterson},
  {Poindexter}, {Sakata}, {Schlesinger}, {Sergeev}, {Skolski}, {Stieglitz},
  {Tobin}, {Unterborn}, {Vestergaard}, {Watkins}, {Watson}, \&
  {Yoshii}}]{2009ApJ...704L..80D}
{Denney}, K.~D., {et~al.} 2009, \apjl, 704, L80

\bibitem[{{Denney} {et~al.}(2010){Denney}, {Peterson}, {Pogge}, {Adair},
  {Atlee}, {Au-Yong}, {Bentz}, {Bird}, {Brokofsky}, {Chisholm}, {Comins},
  {Dietrich}, {Doroshenko}, {Eastman}, {Efimov}, {Ewald}, {Ferbey}, {Gaskell},
  {Hedrick}, {Jackson}, {Klimanov}, {Klimek}, {Kruse}, {Lad{\'e}route}, {Lamb},
  {Leighly}, {Minezaki}, {Nazarov}, {Onken}, {Petersen}, {Peterson},
  {Poindexter}, {Sakata}, {Schlesinger}, {Sergeev}, {Skolski}, {Stieglitz},
  {Tobin}, {Unterborn}, {Vestergaard}, {Watkins}, {Watson}, \&
  {Yoshii}}]{2010ApJ...721..715D}
---. 2010, \apj, 721, 715

\bibitem[{{Dietrich} {et~al.}(1993){Dietrich}, {Kollatschny}, {Peterson},
  {Bechtold}, {Bertram}, {Bochkarev}, {Boroson}, {Carone}, {Elvis},
  {Filippenko}, {Gaskell}, {Huchra}, {Hutchings}, {Koratkar}, {Korista},
  {Lame}, {Laor}, {MacAlpine}, {Malkan}, {Mendes de Oliveira}, {Netzer},
  {Penfold}, {Penston}, {Perez}, {Pogge}, {Richmond}, {Rosenblatt},
  {Shapovalova}, {Shields}, {Smith}, {Smith}, {Sun}, {Thiele}, {Veilleux},
  {Wagner}, {Wilkes}, {Wills}, \& {Wills}}]{1993ApJ...408..416D}
{Dietrich}, M., {et~al.} 1993, \apj, 408, 416

\bibitem[{{Elvis} \& {Karovska}(2002)}]{2002ApJ...581L..67E}
{Elvis}, M., \& {Karovska}, M. 2002, \apjl, 581, L67

\bibitem[{{Horne}(1994)}]{1994ASPC...69...23H}
{Horne}, K. 1994, in Astronomical Society of the Pacific Conference Series,
  Vol.~69, Reverberation Mapping of the Broad-Line Region in Active Galactic
  Nuclei, ed. {P.~M.~Gondhalekar, K.~Horne, \& B.~M.~Peterson}, 23--25

\bibitem[{{Kaspi} {et~al.}(2007){Kaspi}, {Brandt}, {Maoz}, {Netzer},
  {Schneider}, \& {Shemmer}}]{2007ApJ...659..997K}
{Kaspi}, S., {Brandt}, W.~N., {Maoz}, D., {Netzer}, H., {Schneider}, D.~P., \&
  {Shemmer}, O. 2007, \apj, 659, 997

\bibitem[{{Kaspi} {et~al.}(2005){Kaspi}, {Maoz}, {Netzer}, {Peterson},
  {Vestergaard}, \& {Jannuzi}}]{2005ApJ...629...61K}
{Kaspi}, S., {Maoz}, D., {Netzer}, H., {Peterson}, B.~M., {Vestergaard}, M., \&
  {Jannuzi}, B.~T. 2005, \apj, 629, 61

\bibitem[{{Kaspi} {et~al.}(2000){Kaspi}, {Smith}, {Netzer}, {Maoz}, {Jannuzi},
  \& {Giveon}}]{2000ApJ...533..631K}
{Kaspi}, S., {Smith}, P.~S., {Netzer}, H., {Maoz}, D., {Jannuzi}, B.~T., \&
  {Giveon}, U. 2000, \apj, 533, 631

\bibitem[{{Kessler} {et~al.}(2009){Kessler}, {Becker}, {Cinabro}, {Vanderplas},
  {Frieman}, {Marriner}, {Davis}, {Dilday}, {Holtzman}, {Jha}, {Lampeitl},
  {Sako}, {Smith}, {Zheng}, {Nichol}, {Bassett}, {Bender}, {Depoy}, {Doi},
  {Elson}, {Filippenko}, {Foley}, {Garnavich}, {Hopp}, {Ihara}, {Ketzeback},
  {Kollatschny}, {Konishi}, {Marshall}, {McMillan}, {Miknaitis}, {Morokuma},
  {M{\"o}rtsell}, {Pan}, {Prieto}, {Richmond}, {Riess}, {Romani}, {Schneider},
  {Sollerman}, {Takanashi}, {Tokita}, {van der Heyden}, {Wheeler}, {Yasuda}, \&
  {York}}]{2009ApJS..185...32K}
{Kessler}, R., {et~al.} 2009, \apjs, 185, 32

\bibitem[{{Komatsu} {et~al.}(2011){Komatsu}, {Smith}, {Dunkley}, {Bennett},
  {Gold}, {Hinshaw}, {Jarosik}, {Larson}, {Nolta}, {Page}, {Spergel},
  {Halpern}, {Hill}, {Kogut}, {Limon}, {Meyer}, {Odegard}, {Tucker}, {Weiland},
  {Wollack}, \& {Wright}}]{2011ApJS..192...18K}
{Komatsu}, E., {et~al.} 2011, \apjs, 192, 18

\bibitem[{{Krisciunas} {et~al.}(2004){Krisciunas}, {Suntzeff}, {Phillips},
  {Candia}, {Prieto}, {Antezana}, {Chassagne}, {Chen}, {Dickinson},
  {Eisenhardt}, {Espinoza}, {Garnavich}, {Gonz{\'a}lez}, {Harrison}, {Hamuy},
  {Ivanov}, {Krzemi{\'n}ski}, {Kulesa}, {McCarthy}, {Moro-Mart{\'{\i}}n},
  {Muena}, {Noriega-Crespo}, {Persson}, {Pinto}, {Roth}, {Rubenstein},
  {Stanford}, {Stringfellow}, {Zapata}, {Porter}, \&
  {Wischnjewsky}}]{2004AJ....128.3034K}
{Krisciunas}, K., {et~al.} 2004, \aj, 128, 3034

\bibitem[{{Marziani} {et~al.}(2003){Marziani}, {Sulentic}, {Zamanov},
  {Calvani}, {Della Valle}, {Stirpe}, \&
  {Dultzin-Hacyan}}]{2003MSAIS...3..218M}
{Marziani}, P., {Sulentic}, J.~W., {Zamanov}, R., {Calvani}, M., {Della Valle},
  M., {Stirpe}, G., \& {Dultzin-Hacyan}, D. 2003, Memorie della Societa
  Astronomica Italiana Supplementi, 3, 218

\bibitem[{{McKee} \& {Tarter}(1975)}]{1975ApJ...202..306M}
{McKee}, C.~F., \& {Tarter}, C.~B. 1975, \apj, 202, 306

\bibitem[{{Munari} \& {Zwitter}(1997)}]{1997A&A...318..269M}
{Munari}, U., \& {Zwitter}, T. 1997, \aap, 318, 269

\bibitem[{{Perlmutter} {et~al.}(1999){Perlmutter}, {Aldering}, {Goldhaber},
  {Knop}, {Nugent}, {Castro}, {Deustua}, {Fabbro}, {Goobar}, {Groom}, {Hook},
  {Kim}, {Kim}, {Lee}, {Nunes}, {Pain}, {Pennypacker}, {Quimby}, {Lidman},
  {Ellis}, {Irwin}, {McMahon}, {Ruiz-Lapuente}, {Walton}, {Schaefer}, {Boyle},
  {Filippenko}, {Matheson}, {Fruchter}, {Panagia}, {Newberg}, {Couch}, \& {The
  Supernova Cosmology Project}}]{1999ApJ...517..565P}
{Perlmutter}, S., {et~al.} 1999, \apj, 517, 565

\bibitem[{{Peterson}(1993)}]{1993PASP..105..247P}
{Peterson}, B.~M. 1993, \pasp, 105, 247

\bibitem[{{Peterson}(2010)}]{2010IAUS..267..151P}
{Peterson}, B.~M. 2010, in IAU Symposium, Vol. 267, IAU Symposium, 151--160

\bibitem[{{Phillips}(1993)}]{1993ApJ...413L.105P}
{Phillips}, M.~M. 1993, \apjl, 413, L105

\bibitem[{{Riess} {et~al.}(1998){Riess}, {Filippenko}, {Challis},
  {Clocchiatti}, {Diercks}, {Garnavich}, {Gilliland}, {Hogan}, {Jha},
  {Kirshner}, {Leibundgut}, {Phillips}, {Reiss}, {Schmidt}, {Schommer},
  {Smith}, {Spyromilio}, {Stubbs}, {Suntzeff}, \&
  {Tonry}}]{1998AJ....116.1009R}
{Riess}, A.~G., {et~al.} 1998, \aj, 116, 1009

\bibitem[{{Riess} {et~al.}(2001){Riess}, {Nugent}, {Gilliland}, {Schmidt},
  {Tonry}, {Dickinson}, {Thompson}, {Budav{\'a}ri}, {Casertano}, {Evans},
  {Filippenko}, {Livio}, {Sanders}, {Shapley}, {Spinrad}, {Steidel}, {Stern},
  {Surace}, \& {Veilleux}}]{2001ApJ...560...49R}
---. 2001, \apj, 560, 49

\bibitem[{{Schlafly} {et~al.}(2010){Schlafly}, {Finkbeiner}, {Schlegel},
  {Juri{\'c}}, {Ivezi{\'c}}, {Gibson}, {Knapp}, \&
  {Weaver}}]{2010ApJ...725.1175S}
{Schlafly}, E.~F., {Finkbeiner}, D.~P., {Schlegel}, D.~J., {Juri{\'c}}, M.,
  {Ivezi{\'c}}, {\v Z}., {Gibson}, R.~R., {Knapp}, G.~R., \& {Weaver}, B.~A.
  2010, \apj, 725, 1175

\bibitem[{{Schlegel} {et~al.}(1998){Schlegel}, {Finkbeiner}, \&
  {Davis}}]{1998ApJ...500..525S}
{Schlegel}, D.~J., {Finkbeiner}, D.~P., \& {Davis}, M. 1998, ApJ, 500, 525

\bibitem[{{Tonry} {et~al.}(2001){Tonry}, {Dressler}, {Blakeslee}, {Ajhar},
  {Fletcher}, {Luppino}, {Metzger}, \& {Moore}}]{2001ApJ...546..681T}
{Tonry}, J.~L., {Dressler}, A., {Blakeslee}, J.~P., {Ajhar}, E.~A., {Fletcher},
  A.~B., {Luppino}, G.~A., {Metzger}, M.~R., \& {Moore}, C.~B. 2001, \apj, 546,
  681

\bibitem[{{Zu} {et~al.}(2011){Zu}, {Kochanek}, \&
  {Peterson}}]{2011ApJ...735...80Z}
{Zu}, Y., {Kochanek}, C.~S., \& {Peterson}, B.~M. 2011, \apj, 735, 80

\end{thebibliography}
\end{document}